\documentclass[groupedaddress,superscriptaddress,aps,prl,reprint]{revtex4-1}

\usepackage{graphicx}
\usepackage{bm}
\usepackage[normalem]{ulem}
\usepackage{hyperref}
\hypersetup{
setpagesize=false,
 bookmarksnumbered=true,%
 bookmarksopen=true,%
 colorlinks=true,%
 linkcolor=black,
 citecolor=black,
 urlcolor=black,
}
\usepackage{lineno}

\begin{document}

\title{Visualizing half-metallic bulk band structure \\ with multiple Weyl cones of the Heusler ferromagnet}

\author{Takashi~Kono}
\email{takashi-kono@hiroshima-u.ac.jp}
\affiliation{Department of Physical Sciences, Graduate School of Science, Hiroshima University, 1-3-1 Kagamiyama, Higashi-hiroshima 739-8526, Japan}

\author{Masaaki~Kakoki}
\affiliation{Department of Physical Sciences, Graduate School of Science, Hiroshima University, 1-3-1 Kagamiyama, Higashi-hiroshima 739-8526, Japan}

\author{Tomoki~Yoshikawa}
\affiliation{Department of Physical Sciences, Graduate School of Science, Hiroshima University, 1-3-1 Kagamiyama, Higashi-hiroshima 739-8526, Japan}

\author{Xiaoxiao~Wang}
\affiliation{Department of Physical Sciences, Graduate School of Science, Hiroshima University, 1-3-1 Kagamiyama, Higashi-hiroshima 739-8526, Japan}

\author{Kazuki~Goto}
\affiliation{National Institute for Materials Science, 1-2-1 Sengen, Tsukuba 305-0047, Japan}

\author{Takayuki~Muro}
\affiliation{Japan Synchrotron Radiation Research Institute (JASRI), 1-1-1 Kouto, Sayo, Hyogo 679-5198, Japan}

\author{Rie~Y.~Umetsu}
\affiliation{Institute for Materials Research, Tohoku University, 2-1-1 Katahira, Aoba-ku, Sendai 980-8577, Japan}
\affiliation{Center for Spintronics Research Network, Tohoku University, 2-1-1 Katahira, Sendai 980-8577, Japan}
\affiliation{Center for Science and Innovation in Spintronics, Tohoku University, 2-1-1 Katahira, Aoba-ku, Sendai 980-8577, Japan}

\author{Akio~Kimura}
\email{akiok@hiroshima-u.ac.jp}
\affiliation{Department of Physical Sciences, Graduate School of Science, Hiroshima University, 1-3-1 Kagamiyama, Higashi-hiroshima 739-8526, Japan}
\affiliation{Graduate School of Advanced Science and Engineering, Hiroshima University, 1-3-1 Kagamiyama, Higashi-hiroshima 739-8526, Japan}

\date{\today}

\begin{abstract}
Using a well-focused soft X-ray synchrotron radiation beam, angle-resolved photoelectron spectroscopy was applied to a full-Heusler-type Co$_2$MnGe alloy to elucidate its bulk band structure.
A large parabolic band at the Brillouin zone center and several bands that cross the Fermi level near the Brillouin zone boundary were identified in line with the results from first-principles calculations.
These Fermi level crossings are ascribed to majority spin bands that are responsible for electron transport with extremely high spin polarization especially along the direction being perpendicular to the interface of magneto-resistive devices.
The spectroscopy confirms there is no contribution of the minority spin bands to the Fermi surface, signifying half-metallicity for the alloy.
Furthermore, two topological Weyl cones with band crossing points were identified around the $X$ point, yielding the conclusion that Co$_2$MnGe could exhibit topologically meaningful behavior such as large anomalous Hall and Nernst effects driven by the Berry flux in its half-metallic band structure.
\end{abstract}

\maketitle

In general, based on their electronic band structures, solids may be classified as either metal or insulator/semiconductor.
These two different classes may combine in single magnetic crystals, the so-called half-metallic magnets, in which one spin part of the band structure is metallic and the other semiconducting.
Half-metallic magnets have promising uses as spintronics devices because a 100\% spin polarization is expected at the Fermi level.
From first-principle calculations, some of the Heusler alloys have been predicted to possess half-metallic band structures~\cite{Groot1983,Kubler1983,Ishida1995,Picozzi2002}.
Among them, Co$_2$MnSi and Co$_2$MnGe are prototypes that are predicted to exhibit a relatively large minority spin gap, which is preserved as long as they are in the ordered phase.
Later, it was theoretically pointed out that in-gap minority spin states appear when Co antisite defects are created at Mn sites, resulting in much degraded spin polarizations~\cite{Picozzi2004}.
Indeed, Co$_2$MnSi with excess Mn, which prevented Co atoms from occupying Mn sites, had a substantially improved tunneling magnetoresistance ratio~\cite{Ishikawa2009,Liu2012}.
Incorporating Fe atoms at the original Mn sites placed the Fermi level in the center of the minority spin gap and further enhanced the magnetoresistance that relies on the spin polarization by up to 2610\% at 4.2 K~\cite{Moges2016}.
From recent experimental studies, another striking event happens when the Ge site is substituted with Ga.
A huge anomalous Nernst effect takes place~\cite{Sakai2018,Guin2019,Hu2020} caused by a high Berry flux that originates from the bulk band crossings located near the Fermi level ($E_\mathrm{F}$)~\cite{Xiao2006,Xiao2010}.

We thus claim that the locus of $E_\mathrm{F}$ is quite important when engineering the bulk and interface band structures by controlling defects and tuning.
The process of computational material design and experimental confirmation has to be iterated to realize the best materials with extremely high functionality.
To confirm their highly spin-polarized conducting electrons, numerous experiments employing for example point-contact Andreev reflection spectroscopy~\cite{Ritchie2003} and spin-resolved photoelectron spectroscopy~\cite{Fetzer2013,Jourdan2014,Andrieu2016,Guillemard2019_1,Guillemard2019_2} were performed.
However, the three-dimensional nature of this crystal family has prevented us from approaching their bulk band structures mainly because of the surface and interface sensitivities of these techniques.
We note that truly bulk-sensitive hard X-ray photoelectron spectroscopy actually helps in studying the valence band density of states (DOS) of bulk and buried interface, although no momentum-resolved information that is a key to understanding the physical properties has ever been obtained~\cite{Brown1998,Miyamoto2009,Ouardi2011,Kozina2014}.

\begin{figure}
	\centering
	\includegraphics[width=\columnwidth]{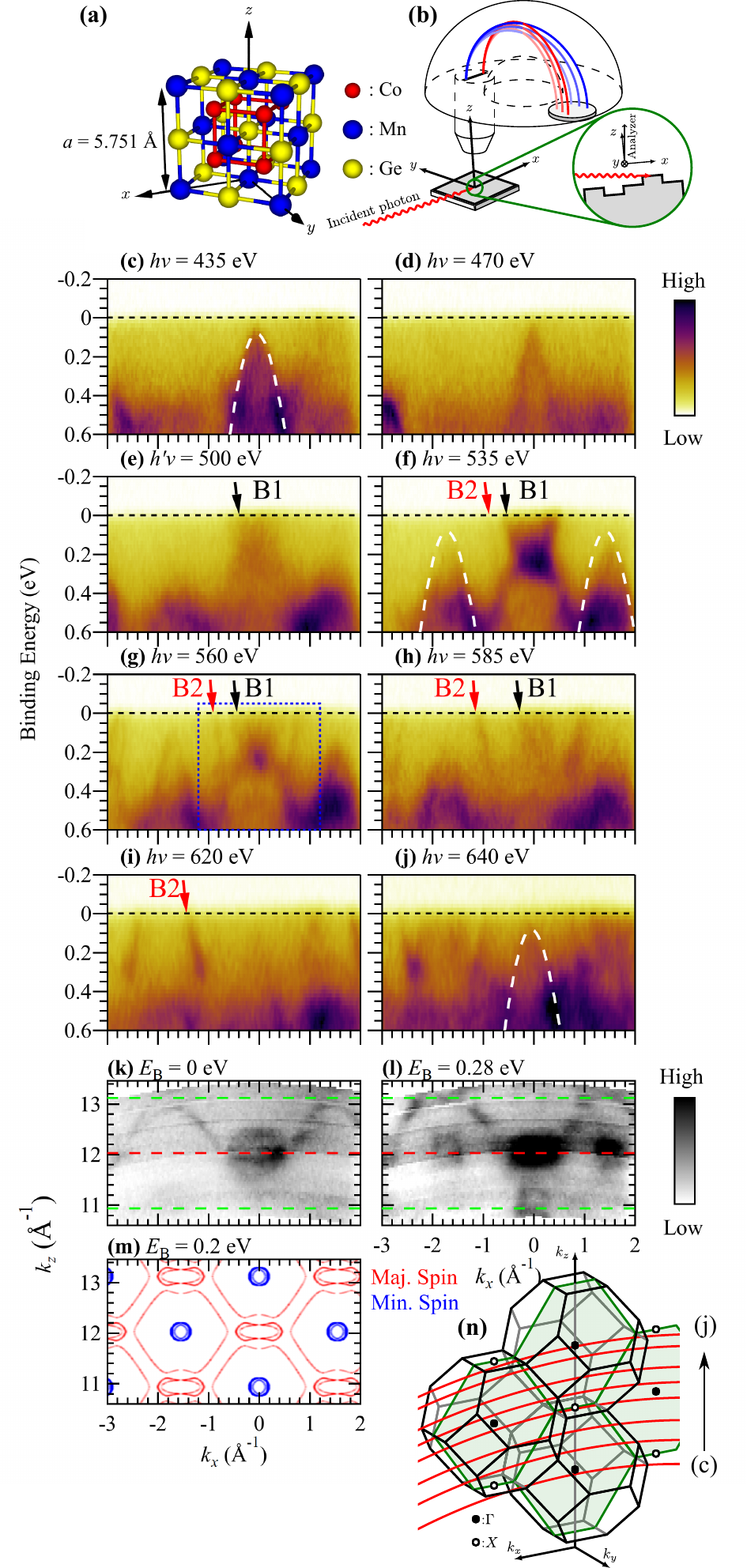}
	\caption{
		(a) Crystal structure of Co$_2$MnGe.
		(b) Experimental setup.
		(c-j) ARPES images v.s.~photon energy.
		(k,~l) Constant energy surface at (k) $E_\mathrm{B}=0$ eV ($E_\mathrm{F}$) and (l) $E_\mathrm{B}=0.28$~eV in $k_x$-$k_z$ plane from ARPES denoted with high symmetry lines.
		Green (red) lines correspond to $k_z=0~(2\pi/a)$ at $k_x=0$.
		(m) Calculated constant energy surfaces ($E_\mathrm{B}=0.2$~eV).
		(n) BZ and $k_x$-$k_z$ surface (shaded by green color).
		Red curves correspond to $k_z$ line of each ARPES images at fixed photon energies.
	}
	\label{fig:kz}
\end{figure}

A recent soft X-ray angle-resolved photoelectron spectroscopy (ARPES) study of Co$_2$MnSi films, each with Al-O$_x$ capping layer, probed mainly the surface and/or interface states, blocking the band structure underneath~\cite{Lidig2019}.
Therefore, to suppress the influence of the results from the surface and/or interface states, it is mandatory to prepare a clean surface by cleaving the crystal under an ultrahigh vacuum.
However, because of the robust 3D nature of its crystal structure [Fig.~\ref{fig:kz}(a)], it is generally difficult to produce a flat surface and consequently multiple steps remain [Fig.~\ref{fig:kz}(b)].

In this \textit{Letter},  we report the direct evidence of the half-metallic band structure of Co$_2$MnGe and the presence of multiple Weyl cones using ARPES with a beam spot size of $10 \times 10~\mu$m.
It enables a single domain with a flat surface to be proved for ARPES\@.

A high quality single crystalline sample was grown by the Bridgeman method (see Sec.~S1 of Supplementary Material for the detailed sample growth conditions).
The composition of the sample was Co: 49.3, Mn: 24.9, and Ge: 25.8 (at.~\%) as determined by energy dispersive X-ray microanalysis (EDX), which is stoichiometric enough to prevent Co anti-site defect and preserve its half-metallicity.
The saturation magnetic moment evaluated from the magnetization curve measured with a SQUID at 5~K was 115.3~emu/g and converted to 4.96~$\mu_\mathrm{B}/\mathrm{f.u.}$, being close to an integer number and consistent with the expected value obtained from the Slater-Pauling rule~\cite{Galanakis2002}.
From differential scanning calorimetry, the Curie temperature of the present specimen was 913~K, being comparable with our previous experimental result~\cite{Okubo2010}.

ARPES measurements were performed at BL25SU at SPring-8.
Circularly polarized synchrotron radiation beam was used to maximize the number of bands that appear in the ARPES results.
The beam spot size was adjusted to $10 \times 10~\mu$m.
The energy and angular resolutions for ARPES were set to $<80$~meV and $<0.2^\circ$, respectively.
A clean $(001)$ surface of Co$_2$MnGe was obtained by cleaving in an ultra-high vacuum (pressure $<2\times 10^{-8}$~Pa).
All measurements were conducted at a temperature of $\sim30$~K.
The analyzer slit was set along the $k_x$-axis [parallel to $[110]$; see Figs.~\ref{fig:kz}(a,~b)].

All first-principles density-functional calculations were performed using the WIEN2k program code~\cite{wien2k}.
We used the spin-polarized generalized gradient approximation~\cite{Perdew1997} (see Sec.~S2 for further calculation details).
The Coulomb interaction $U$ was not considered because previous studies indicated that $U$ plays a minor role in Co$_2$MnGe~\cite{Tsirogiannis2015,Sharma2016}.

From the photon energy ($h\nu$) dependence of the ARPES spectra [Figs.~\ref{fig:kz}(c-j)], we see that the ARPES image evolves with photon energy (also see the Supplemental Movie).
At $h\nu=435$~eV, a band that disperses downwards from its peak at $E_\mathrm{B}=0.1$~eV appears at $k_x=0$ [overlaid with white dashed line in Fig.~\ref{fig:kz}(c)].
The photoelectron intensity from this band weakens at $h\nu=470$~eV [Fig.~\ref{fig:kz}(d)], and another feature emerges at $h\nu=500$~eV [B1 in Fig.~\ref{fig:kz}(e)] and its intensity is maximized at $h\nu=535$~eV [Fig.~\ref{fig:kz}(f)].
Interestingly, the bands cross $E_\mathrm{F}$ (B1, B2) and two bands, each with a downward dispersion, emerge on both sides of the $E_\mathrm{F}$ crossing feature (white dashed lines).
Increasing the photon energy further, the band indicated by the black arrow moves to $k_x=0~\text{\AA}^{-1}$.
Moreover, a steeply dispersing band crosses $E_\mathrm{F}$ [red arrow labeled with B2 in Fig.~\ref{fig:kz}(f)] and moves away from $k_x=0~\text{\AA}^{-1}$ with higher photon energies.
A spectral weight of the band, indicated by the red arrow, is maximized at $h\nu=620$~eV [Fig.~\ref{fig:kz}(i)].
The downward-dispersing band re-emerges above $h\nu=620$~eV, and its intensity is maximized at $h\nu=640$~eV [dashed line in Fig.~\ref{fig:kz}(j)].

We realized that the circular features in the $k_x$-$k_z$ constant energy map at $E_\mathrm{B}=0.28$~eV [Fig.~\ref{fig:kz}(l)] stem from the downward-dispersing bands near the $\Gamma$ point.
Importantly, they are not seen at $E_\mathrm{F}$ [Fig.~\ref{fig:kz}(k)].
This demonstrates that our ARPES measurement tracks the band structure along the out-of-plane momentum ($k_z$) line.
We have determined an inner potential of $V_0=18$~eV from the photon energy dependence of the ARPES spectra with downward dispersions.
The computed $k_x$-$k_z$ constant energy map [Fig.~\ref{fig:kz}(m)] reproduces well the observed features with minority spin character.

\begin{figure*}
	\centering
	\includegraphics[width=0.8\textwidth]{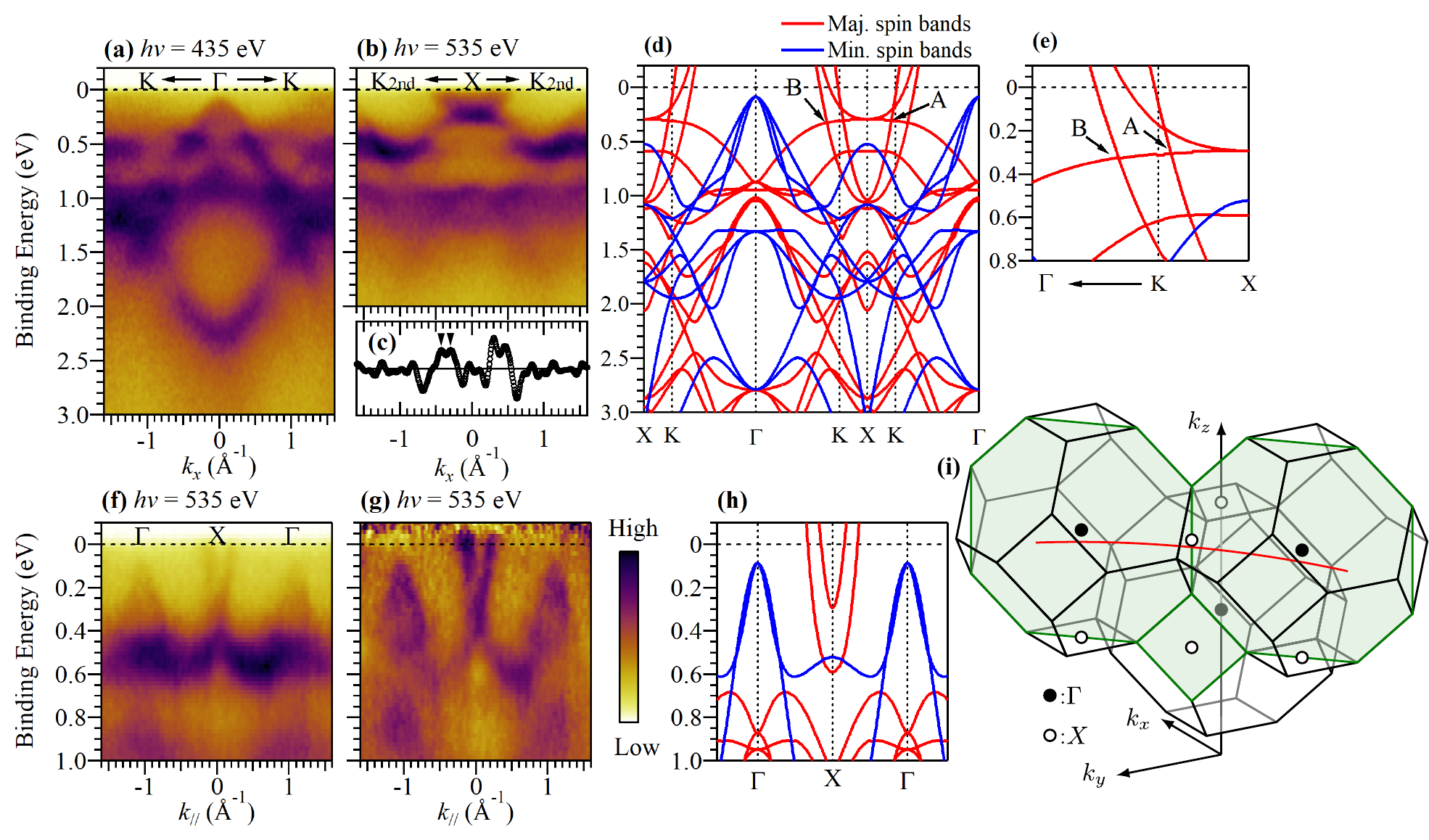}
	\caption{
		(a,~b) ARPES images measured with a photon energies ($h\nu$s) of 435 and 535~eV [same $k$ cuts with Figs.~\ref{fig:kz}(c,~f), respectively].
		(c) Second derivative momentum distribution curve along $E_\mathrm{B}=0$~eV line ($E_\mathrm{F}$) in panel (b).
		(d) Calculated band dispersions along $\Gamma$-$K$-$X_\mathrm{2nd}$ line.
		Red (blue) color corresponds to majority (minority) spin components.
		(e) Calculated band dispersions around $K$ and $X$ point.
		(f) Measured ARPES band dispersions along $\Gamma$-$X$ line. [Red dashed line in Fig.~\ref{fig:kxky}(c)]
		(g) ARPES image normalized by integrated intensity of momentum distribution curve from panel (f).
		(h) Calculated band dispersions along $\Gamma$-$X$ line.
		(i) BZ and high symmetry points.
		Red line corresponds to $k_z$ line of panel (f).
	}
	\label{fig:BD}
\end{figure*}
We extracted typical ARPES images acquired at the incident photon energies of 435~eV and 535~eV from Fig.~\ref{fig:kz} along the $K$-$\Gamma$-$K$ and $K_\mathrm{2nd}$-$X$-$K_\mathrm{2nd}$ lines [Figs.~\ref{fig:BD}(a,~b), respectively].
A parabolic band with a downward dispersion dominates near $k_x=0~\text{\AA}^{-1}$ at $h\nu=435$~eV\@.
At $h\nu=535$~eV, a spectral feature near $E_\mathrm{F}$ around $k_x=0~\text{\AA}^{-1}$ is modified considerably.
This feature corresponds to B1 in Fig.~\ref{fig:kz} and consists of two separated bands that are more clearly identified by the second derivative of the momentum distribution curve at $E_\mathrm{F}$ [indicated by inverted triangles in Fig.~\ref{fig:BD}(c)].
By combining this plot with B2 in Fig.~\ref{fig:kz}, we conclude that the three bands cross $E_\mathrm{F}$.
Furthermore, we compared these ARPES images with calculated band dispersions along the $\Gamma$-$K$-$X_\mathrm{2nd}$ line [Fig.~\ref{fig:BD}(d)].
The calculated results have been shifted in energy by $-100$~meV (40~meV) toward higher $E_\mathrm{B}$ in the majority (minority) spin channel to adjust to the ARPES results.
The downward-dispersing ARPES band [Fig.~\ref{fig:BD}(a)] fits very well with the theoretical minority spin band.
A band that disperses upwards and crosses $E_\mathrm{F}$ [Fig.~\ref{fig:BD}(b)] is ascribed to the three computed majority spin bands.

For the experimental band dispersion along the $\Gamma$-$X$ line [Fig.~\ref{fig:BD}(f)], the raw ARPES intensity has been normalized to enhance its visibility using the integrated intensity of the momentum distribution curve [Fig.~\ref{fig:BD}(g)].
From a comparison with the calculated band dispersion along the $\Gamma$-$X$ line [Fig.~\ref{fig:BD}(h)], we claim that the minority and majority spin components contribute respectively to the observed bands that disperse downwards near the $\Gamma$ points and the band that disperses upwards and crosses $E_\mathrm{F}$.

\begin{figure*}
	\centering
	\includegraphics[width=0.8\textwidth]{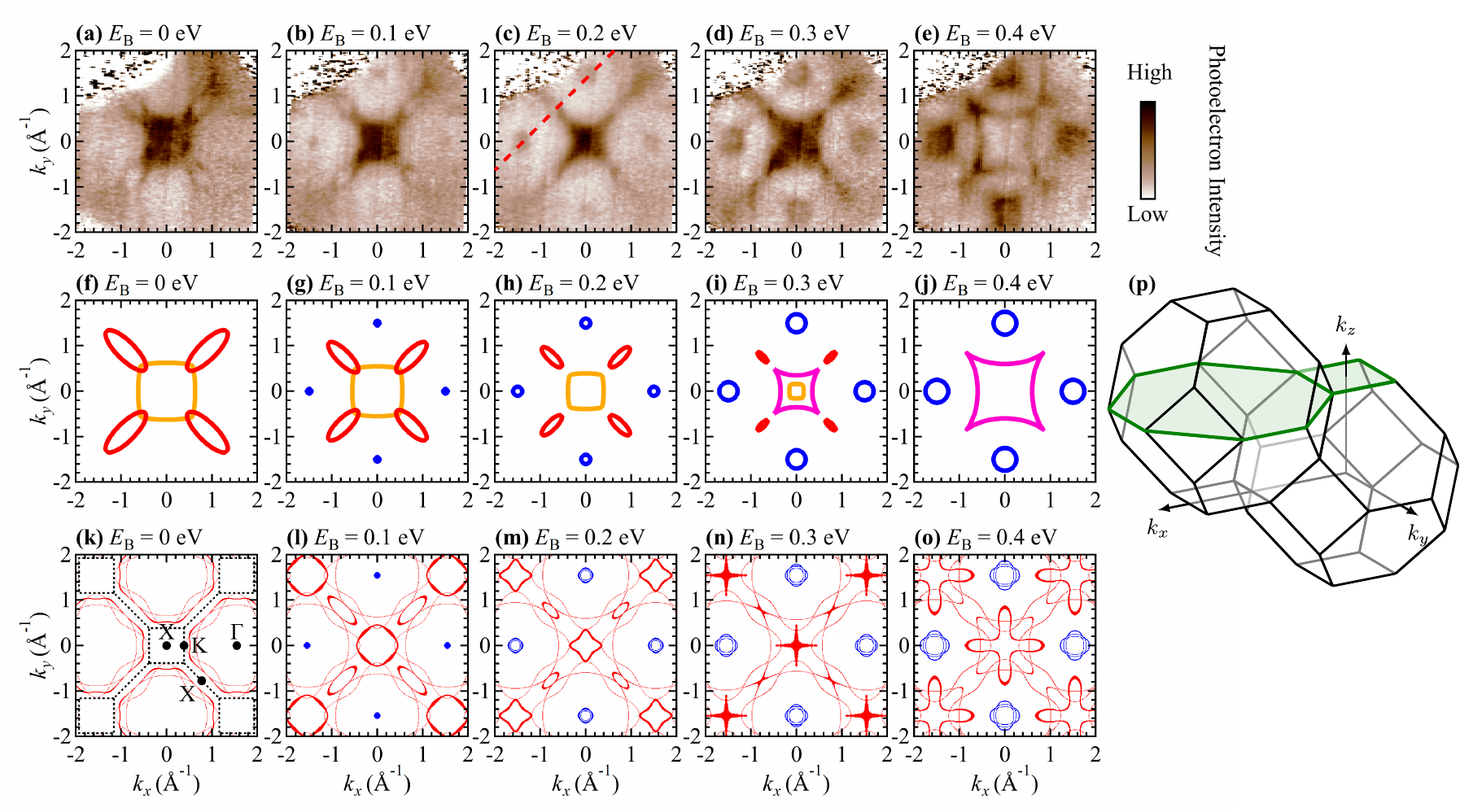}
	\caption{
		(a-e) Constant energy surfaces in $k_x$-$k_y$ plane mapped by ARPES with photon energy of 535~eV\@.
		(f-j) Characteristic features of constant energy surfaces depicted from panels (a-e).
		(k-o) Calculated constant energy surfaces with an energy offset of $-100$~meV (40~meV) toward higher $E_\mathrm{B}$ in a majority (minority) spin channel.
		(p) BZ and $k_x$-$k_y$ surface (shaded by green color).
	}
	\label{fig:kxky}
\end{figure*}
Figures~\ref{fig:kxky}(a-e) shows the experimental constant energy surfaces at $E_\mathrm{B}=0$, 0.1, 0.2, 0.3 and 0.4~eV\@.
The characteristic features in Figs.~\ref{fig:kxky}(a-e) are re-drawn in Figs.~\ref{fig:kxky}(f-j).
The calculated constant energy surfaces are shown in Figs.~\ref{fig:kxky}(k-o).
Constant energy surfaces with four-fold symmetry are observed that correlate with the symmetry of the $(001)$ plane of the crystal.
Four elliptical pockets [red solid line in Fig.~\ref{fig:kxky}(f)] are seen at $E_\mathrm{B}=0$~eV and diminish as $E_\mathrm{B}$ increases.
At $E_\mathrm{B}=0.4$~eV [Figs.~\ref{fig:kxky}(e,~j)], we find four circles that diminish with decreasing $E_\mathrm{B}$ and ultimately disappear at $E_\mathrm{F}$.
In a comparison with the calculated results [Figs.~\ref{fig:kxky}(k-o)], we determined that these circles are ascribable to the minority spin components.
All the other features are well explained by the bands of the majority spin channel.
This means that only the majority spin bands cross $E_\mathrm{F}$ and no minority spin band exists, signifying that a half-metallic band structure is realized in this crystal.

\begin{figure}
	\centering
	\includegraphics[width=\columnwidth]{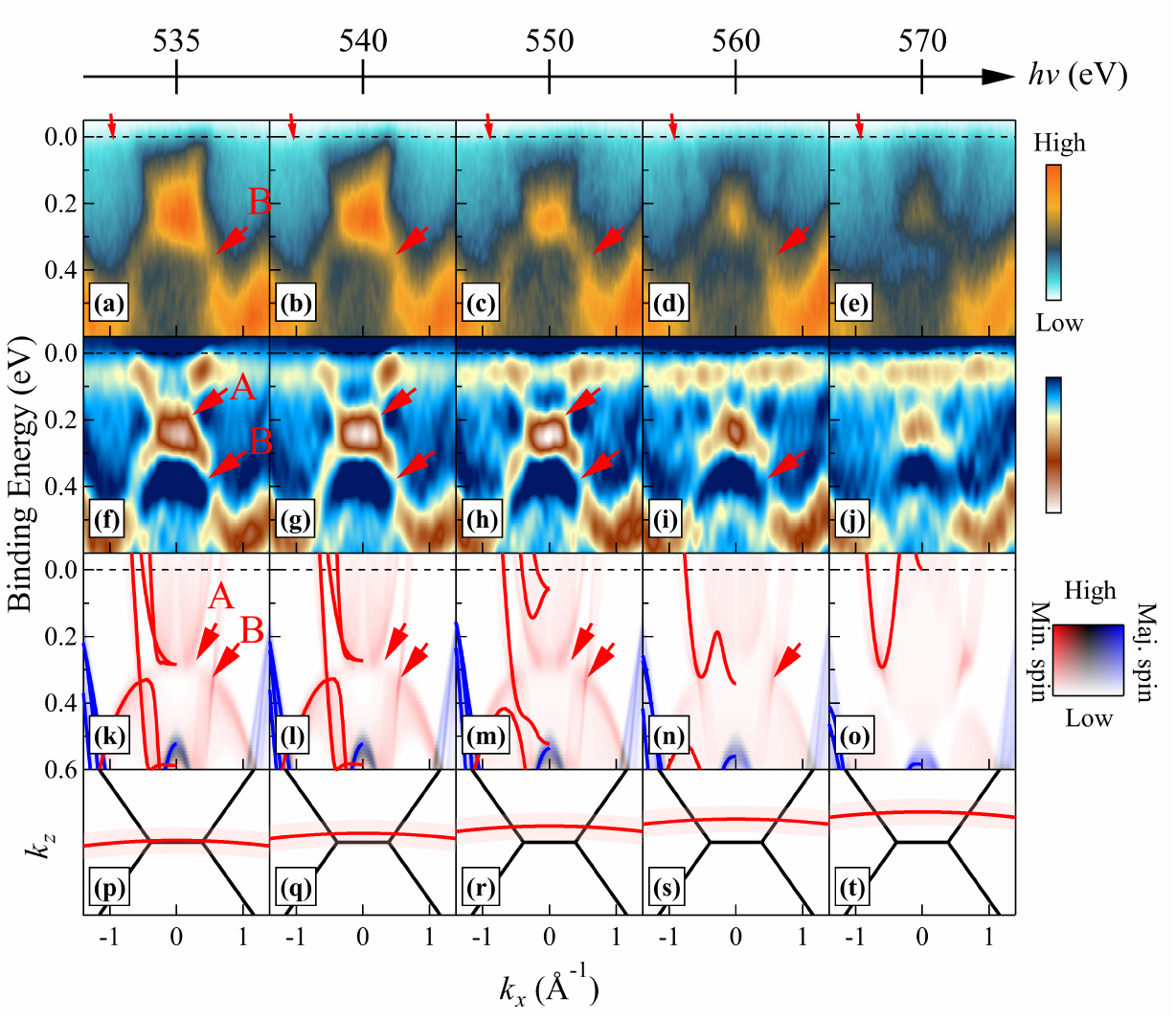}
	\caption{
		(a-e) ARPES images with changing photon energy around $h\nu=535$~eV\@.
		Panel (d) correspond to blue dotted box in Fig.~\ref{fig:kz}(g).
		(f-j) Second derivative ARPES images from panels (a-e).
		(k-o) Calculated band dispersion with an energy offset of $-100$~meV (40~meV) toward higher $E_\mathrm{B}$ in a majority (minority) spin channel.
		Momentum path is simulated from kinetic energy of photoelectron and considering $k_z$ broadening of $\delta k_z \sim \pm 0.2~\text{\AA}^{-1}$.
		(p-t) Red curves are correspond to each ARPES images and calculated results.
	}
	\label{fig:cross}
\end{figure}
Finally, we discuss the topological aspects of the band structure of Co$_2$MnGe.
Weyl fermions appear in materials that break inversion symmetry~\cite{Xu2015} or time-reversal symmetry~\cite{Kuroda2017}.
It is expected that the Weyl fermions emerge also in Co$_2$MnGe as a result of time-reversal symmetry breaking as has recently been predicted for some of the Heusler alloys~\cite{Chang2017,Belopolski2019}.
Figures~\ref{fig:cross}(a-e) show the ARPES images along the $K_\mathrm{2nd}$-$X$-$K_\mathrm{2nd}$ line acquired at $h\nu=535$-570~eV\@.
To exclude the effect of a momentum constant background, we differentiated the ARPES intensity twice along the energy axis in Figs.~\ref{fig:cross}(f-j).
We see again a band that disperses upwards and crosses $E_\mathrm{F}$, at $h\nu=535$~eV [Figs.~\ref{fig:cross}(a,~f)].
The inner band exhibits a strong intensity whereas the outer band [small red arrow in Fig.~\ref{fig:cross}(a)] is weak.
Marked by two red arrows A and B, the band dispersing downwards intersects other bands dispersing upwards.
These two crossings are persistently seen in the image taken at $h\nu=550$~eV [Fig.~\ref{fig:cross}(h)].
At $h\nu=560$~eV, the crossing of the bands involving the outer band is observed even more clearly at $E_\mathrm{B}=0.4$~eV\@.
Above $h\nu=570$~eV, all crossing points disappear.
To compare the ARPES result with the theoretical band dispersions more rigorously, we considered real momentum ``paths'' using $k_z=\sqrt{2m_e(h\nu-W+V_0)/\hbar^2-k_x^2}~(W=4.5~\text{eV})$.
It is also necessary to consider $k_z$ broadening effects that stem from the limited probing depth of the photoelectrons ($\delta k_z=\pm0.2~\text{\AA}^{-1}$).
In Figs.~\ref{fig:cross}(k-o), we show the computed band dispersions integrated in the $\delta k_z$ window together with the band dispersions along the real $k$-lines [Figs.~\ref{fig:cross}(p-t)].
We find that the theoretical ARPES images reproduce the persistent band crossings for $h\nu=540$-560~eV and its disappearance above $h\nu=570$~eV, although the crossings of the line dispersions are already broken above 550~eV\@.
Our first-principles calculation predicts two band crossings around the $X$ point [A and B in Fig.~\ref{fig:BD}(e)].
Further theoretical analysis tells us that all of them form nodal-lines [Figs.~S3(a-c)].
We, therefore, conclude that crossing points A and B correspond to the type-II Weyl points produced from the tilted cones.
We note that Co$_2$MnGa exhibits similar crossing points [Figs.~S3(d-f)] and that these crossings may generate a high Berry flux~\cite{Chang2017,Belopolski2019}.
The gigantic anomalous electrical and thermal conductivities that give rise to the anomalous Hall effect and the Nernst effect emerge when the gap opens at the crossing point and $E_\mathrm{F}$ is tuned inside the gap.
Because the crossing points A and B are located above $E_\mathrm{F}$ for Co$_2$MnGa, further carrier tuning is required to improve these anomalous conductivities.
We propose that the substitution of Ge atoms into the Ga sites may improve the anomalous transport properties.
Our study with soft X-ray ARPES confirms that these crossing points are of bulk origin and are maintained even for the end material, Co$_2$MnGe. 

In conclusion, we performed ARPES on a full-Heusler-type Co$_2$MnGe bulk crystal utilizing micro-spot-size soft X-ray synchrotron radiation.
No contribution of the minority spin band structures to the Fermi surface was observed.
All the observed Fermi surfaces were reproduced by the calculated results for the majority spin channel.
Moreover, two topological Weyl cones were clearly observed indicating Berry flux sources.
Our findings provide strong evidence of half-metallicity coexisting with multiple Weyl cones of the Co$_2$MnGe alloy.
They also shed light on the currently elusive spin-gapless Heusler semiconductors, for which one spin part is semimetallic and the other semiconducting.
Tuning the highly spin-polarized carriers is possible via an electric field if that becomes feasible.
We finally remark that by choosing appropriate elements the Heusler alloys comprising more than three elements give rise to various types of physical behaviors such as the magneto-caloric effect, thermoelectricity and superconductivity, all of which rely on their band structures.
Micro-spot ARPES with soft X-ray synchrotron radiation beam affords opportunities to realize highly functional materials for such alloys.

\begin{acknowledgements}
This work was financially supported by KAKENHI (Nos.~17H06152, 17H06138, 18H01690, and 18H03683).
The soft X-ray ARPES experiment was performed with the approval of JASRI (Proposal No.~2019A1548).
Micro-ARPES instruments was developed by Photon and Quantum Basic Research Coordinated Development Program from MEXT\@.
T.Y. was financially supported by Grants-in-Aid for JSPS Fellows No.~18J22309.
We sincerely thank T.~Sugawara and I.~Narita in Tohoku University for their help to make the single crystals and to perform the EDX experiment.
\end{acknowledgements}

\bibliography{lib_main_no_url}

\end{document}



\title{Supplemental Materials:\\ Visualizing half-metallic bulk band structure \\ with multiple Weyl cones of the Heusler ferromagnet}

\author{Takashi~Kono}
\email{takashi-kono@hiroshima-u.ac.jp}
\affiliation{Department of Physical Sciences, Graduate School of Science, Hiroshima University, 1-3-1 Kagamiyama, Higashi-hiroshima 739-8526, Japan}

\author{Masaaki~Kakoki}
\affiliation{Department of Physical Sciences, Graduate School of Science, Hiroshima University, 1-3-1 Kagamiyama, Higashi-hiroshima 739-8526, Japan}

\author{Tomoki~Yoshikawa}
\affiliation{Department of Physical Sciences, Graduate School of Science, Hiroshima University, 1-3-1 Kagamiyama, Higashi-hiroshima 739-8526, Japan}

\author{Xiaoxiao~Wang}
\affiliation{Department of Physical Sciences, Graduate School of Science, Hiroshima University, 1-3-1 Kagamiyama, Higashi-hiroshima 739-8526, Japan}

\author{Kazuki~Goto}
\affiliation{National Institute for Materials Science, 1-2-1 Sengen, Tsukuba 305-0047, Japan}

\author{Takayuki~Muro}
\affiliation{Japan Synchrotron Radiation Research Institute (JASRI), 1-1-1 Kouto, Sayo, Hyogo 679-5198, Japan}

\author{Rie~Y.~Umetsu}
\affiliation{Institute for Materials Research, Tohoku University, 2-1-1 Katahira, Aoba-ku, Sendai 980-8577, Japan}
\affiliation{Center for Spintronics Research Network, Tohoku University, 2-1-1 Katahira, Sendai 980-8577, Japan}
\affiliation{Center for Science and Innovation in Spintronics, Tohoku University, 2-1-1 Katahira, Aoba-ku, Sendai 980-8577, Japan}

\author{Akio~Kimura}
\email{akiok@hiroshima-u.ac.jp}
\affiliation{Department of Physical Sciences, Graduate School of Science, Hiroshima University, 1-3-1 Kagamiyama, Higashi-hiroshima 739-8526, Japan}
\affiliation{Graduate School of Advanced Science and Engineering, Hiroshima University, 1-3-1 Kagamiyama, Higashi-hiroshima 739-8526, Japan}

\maketitle

\renewcommand{\thefigure}{S\arabic{figure}}
\renewcommand{\thesection}{S\arabic{section}}
\renewcommand*{\citenumfont}[1]{S#1}
\renewcommand*{\bibnumfmt}[1]{[S#1]}

\section{S\lowercase{ample preparations}}
The mother ingot of the polycrystalline Co$_2$MnGe was fabricated by induction melting in an argon gas atmosphere.
A single crystal with a diameter size of 12~mm and a length of about 30~mm was grown by the Bridgeman method.
The obtained ingot was annealed at 1273~K and slowly cooled down to room temperature.
The crystal orientation was checked using Laue's back-reflection method and the specimen was cut into stripes in the direction parallel to the \verb|<|100\verb|>|.
From energy dispersive X-ray spectroscopy, the sample's composition was evaluated to be Co:~49.3, Mn:~24.9, and Ge:~25.8 (at.~\%).
This is stoichiometric enough to prevent Co anti-site defect and preserve its half-metallicity~[4, 5].

\section{F\lowercase{irst-principle calculation}}
The muffin-tin approximation was used for the potential; the muffin-tin radius $R_\mathrm{MT}$ of each atom was taken to be $R_\mathrm{MT}^\mathrm{Co}=R_\mathrm{MT}^\mathrm{Mn}=2.30$~Bohr, and $R_\mathrm{MT}^\mathrm{Ge}=2.23$~Bohr.
The wave functions were expanded into spherical harmonics with integer $\ell$ ranging up to $\ell_\mathrm{max}=10$ in the muffin-tin spheres and by plane waves in the interstitial region with a cut-off value of $R_\mathrm{MT}^\mathrm{Ge}\cdot K_\mathrm{max}=7$.
The Fourier charge density was expanded up to $G_\mathrm{max}=12~\mathrm{Bohr}^{-1}$.
The $k$ space was divided into a uniform $21\times 21\times 21$ mesh.
These $RK_\mathrm{max}$, $\ell_\mathrm{max}$, $G_\mathrm{max}$ and $k$-points were sufficient to stabilize the shape of the DOS\@. 
In this calculation, we set the lattice constants to $a=b=c=5.751~\text{\AA}$ and $\alpha=\beta=\gamma=90^\circ$ (the experimental values reported in Ref.~[26] of the main text), and the L2$_1$ ($Fm\bar{3}m$) phase with atomic positions of Co~$(0.25,0.25,0.25)$, Mn~$(0,0,0)$, and Ge~$(0.5,0.5,0.5)$.

\section{P\lowercase{hoton energy dependence of} ARPES \lowercase{spectra}}
As shown in Supplemental movie, we can see a quasi-continuous change in the ARPES band structure by changing incident photon energy.
This further clarifies that our ARPES measurement with variable incident photon energy in the soft X-ray region can track the band structure in three-dimension.

Here, we describe how the inner potential is determined from $h\nu$ dependence of ARPES spectra.
The vertical component of the wavenumber of the $N^\mathrm{th}$ $\Gamma$ point is expressed as $k_\perp=4\pi N/a$ using the lattice constant ($a$).
This is due to the vertical periodicity of BZ.
On the other hand, to estimate $k_\perp$ from the ARPES measurements, the information on the photoemission final state is necessary.
However, it is difficult to observe or rigorously calculate the band structure of the final state, so we assume that the final state is a free electron.
The parameter characterizing the free electrons is the inner potential ($V_0$).
Using the photon energy ($h\nu$) where the electronic structure at the $\Gamma$ point (in this case a convex parabolic band) is observed by ARPES measurements, there exists pairs of $N,V_0$ yielding $k_\perp=4\pi N/a=\sqrt{2m_e(h\nu-W+V_0)/\hbar^2}$.
Here, if there exists a natural number $N$ such that $V_0$ takes a reasonable value that fits to the realistic band structure.
In this particular case of Co$_2$MnGe, the valence band bottom is located at 12~eV below the Fermi level~[22] and the obtained value of $V_0~(=18~\mathrm{eV})$ is found to be reasonable when the work-function (energy distance between the Fermi level and the vacuum level) is considered.
Therefore, we can say that the inner potential is not only a fitting parameter but also an important physical quantity to determine whether the final state can be assumed as a free electron or not.

\section{C\lowercase{omputational simulation of nodal-lines}}
\begin{figure*}
	\centering
	\includegraphics[width=\columnwidth]{BD_p.pdf}
	\caption{\textbf{Band crossing points on high-symmetry lines.}
	\textbf{(a,~c)}~Band crossing points of (a)~Co$_2$MnGe and (c)~Co$_2$MnGa.
	\textbf{(b,~d)}~Enlarged view of cross points E and F in panel~(a,~c), respectively.
	\textbf{(e)}~Crystal structure of Co$_2$MnGe.
	}
	\label{fig:s_BD}
\end{figure*}

Figure~\ref{fig:s_BD} shows the theoretical band dispersions along high-symmetry lines of Co$_2$MnGe and Co$_2$MnGa.
We find some band crossing points labeled A to F in the vicinity of the $E_\mathrm{F}$.
Here, Band-1 and Band-2 form crossing points A, E, and F\@.
On the other hand, Band-2 and Band-3 generate crossing points B, C, and D\@.
We confirm that these degenerate points form nodal-lines (NLs) in three-dimensional $k$-space.

\begin{figure*}
	\centering
	\includegraphics{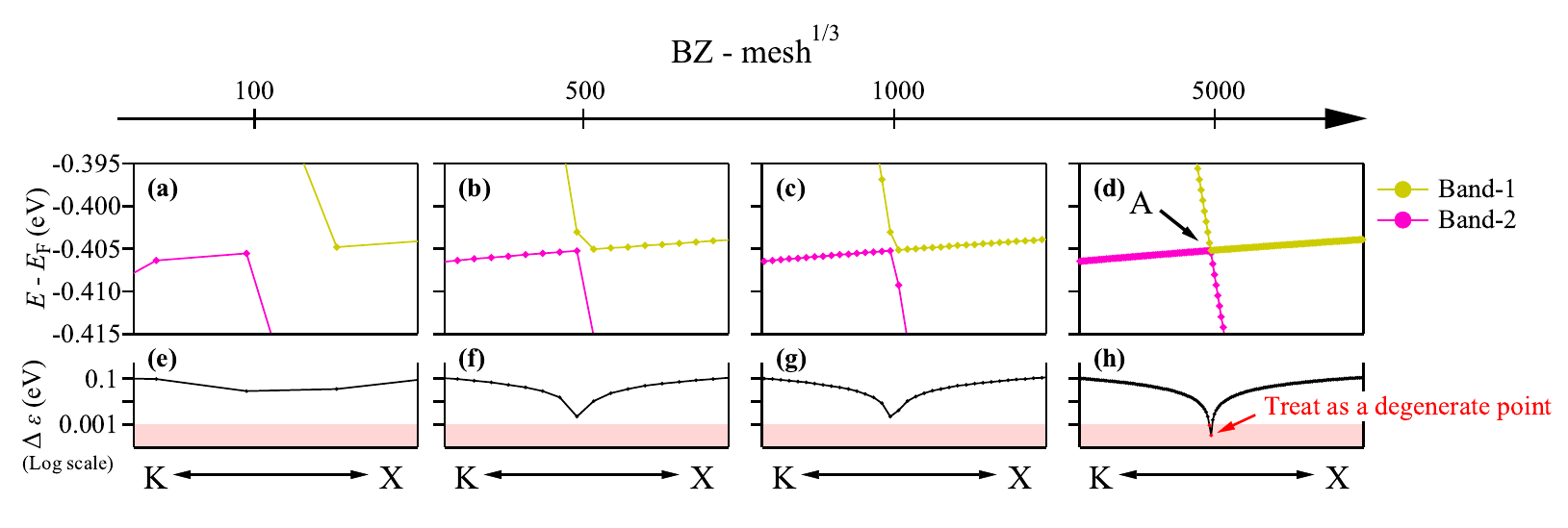}
	\caption{\textbf{BZ-mesh dependence of the band crossing feature near the point A of Co$_2$MnGe in Fig.~\ref{fig:s_BD}(a).}
		\textbf{(a-d)}~Energy eigenvalues of Band-1 and Band-2.
		\textbf{(e-h)}~Size of band gap between Band-1 and Band-2.
	}
	\label{fig:s_BZmesh}
\end{figure*}

\begin{figure*}
	\centering
	\includegraphics{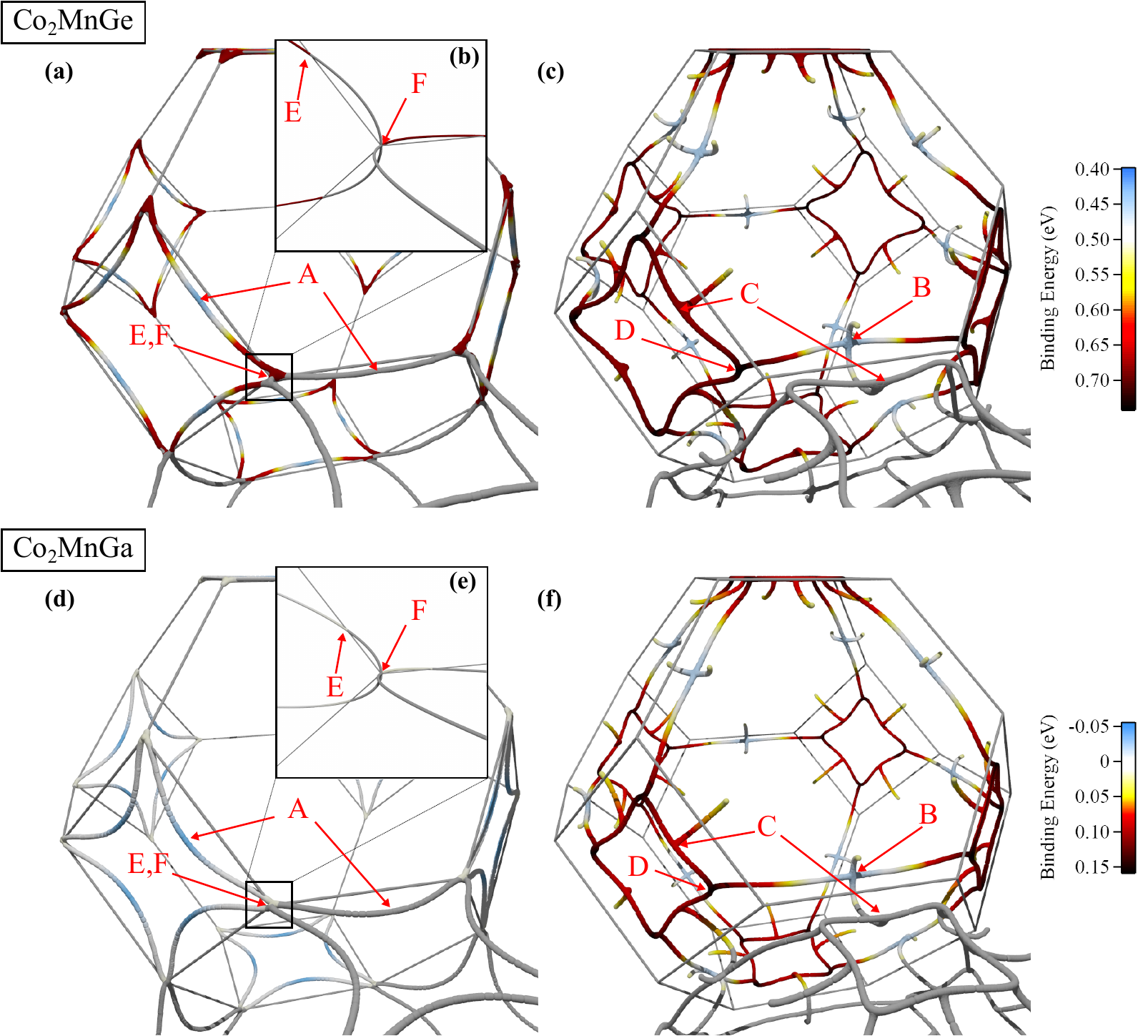}
	\caption{\textbf{Theoretical simulation of NLs.}
	\textbf{(a,~c)}~Degenerate points between (a)~Band-1 and -2 and (c)~Band-2 and -3 of Co$_2$MnGe.
	\textbf{(b)}~Enlarged view around $W$ point in panel~(a).
	\textbf{(d,~f)}~Degenerate points between (d)~Band-1 and -2 and (f)~Band-2 and -3 of Co$_2$MnGa.
	\textbf{(e)}~Enlarged view around $W$ point in panel~(d).
	Color correspond to binding energy of degenerate points.
	Degenerate points in second BZ are shown in gray.
	}
	\label{fig:s_NL}
\end{figure*}

We calculated the size of band gap $(\Delta \varepsilon)$ between Band-1(2) and Band-2(3) in the whole BZ to visualize NLs.
Computational calculation gives us finite values of $\Delta \varepsilon$ even if there is a degenerate point.
$\Delta \varepsilon$ converges to zero with infinite BZ-mesh (the number of $k$ points in the BZ).
Figure~\ref{fig:s_BZmesh} shows BZ-mesh dependence of the crossing point A of Co$_2$MnGe.
Band gaps between Band-1 $(\varepsilon_1)$ and Band-2 $(\varepsilon_2)$ are calculated in each $k$ points with $\Delta \varepsilon (k)=|\varepsilon_1(k)-\varepsilon_2(k)|$, and shown in Figs.~\ref{fig:s_BZmesh}(e-h).
For $100\times100\times100$ mesh, the band dispersion is found to be unclear [Fig.~\ref{fig:s_BZmesh}(a)] and it is difficult to find out degenerate points from Fig.~\ref{fig:s_BZmesh}(e).
The minimum value of $\Delta \varepsilon$ gets smaller as the number of BZ-mesh increases.
For the finest mesh in our study ($5000\times5000\times5000$) used in the main NL calculation, we can see a crossing feature in the band dispersion [Fig.~\ref{fig:s_BZmesh}(d)] and the degenerate point is obvious from $\Delta\varepsilon(k)$ [Fig.~\ref{fig:s_BZmesh}(h)].
In the main NL calculation, we plot the $k$ point as a degenerate point that satisfies $\Delta \varepsilon(k) < 0.001$~eV in Fig.~\ref{fig:s_NL}.

Calculated degenerate points are shown in Fig.~\ref{fig:s_NL}.
We realize that all the degenerate points between Band-1(2) and Band-2(3) form one-dimensional lines.
Crossing points A to F in Fig.~\ref{fig:s_BD} are indicated by red arrow in Fig.~\ref{fig:s_NL}.
Our calculated NLs formed by Band-2 and Band-3 of Co$_2$MnGa [Fig.~\ref{fig:s_NL}(f)] reproduce previous theoretical results~\cite{Chang2017}, and we confirmed that Co$_2$MnGe has similar NLs [Fig.~\ref{fig:s_NL}(c)].
Furthermore, the degenerate points between Band-1 and Band-2 form another NL [Figs.~\ref{fig:s_NL}(a,~d)].
From ARPES, we identified crossing points A and B (Fig.~4) for which both of the corresponding NLs are formed between Bands-1 and -2 and between Bands-2 and -3.

\bibliography{lib_main_no_url}